\documentclass[%
 aip,
 amsmath,amssymb,
 reprint,%
floatfix
]{revtex4-1}

\usepackage{graphicx}% Include figure files
\usepackage{dcolumn}% Align table columns on decimal point
\usepackage{bm}% bold math
\usepackage[utf8]{inputenc}
\usepackage[T1]{fontenc}
\usepackage{mathptmx}
\usepackage{etoolbox}
\usepackage{pgffor}

\makeatletter
\def\@email#1#2{%
 \endgroup
 \patchcmd{\titleblock@produce}
  {\frontmatter@RRAPformat}
  {\frontmatter@RRAPformat{\produce@RRAP{*#1\href{mailto:#2}{#2}}}\frontmatter@RRAPformat}
  {}{}
}%
\makeatother
\begin{document}

\preprint{AIP/123-QED}

\title[MLP highlights melting and freezing of AlNPs]{Machine-learnt potential highlights melting and freezing of aluminium nanoparticles}
% Force line breaks with \\
\author{Davide Alimonti}
\author{Francesca Baletto}%
 \email{francesca.baletto@unimi.it}
\affiliation{ 
Physics Department,  University of Milan, Via Celoria 16, 20133 Milano, Italy
}%

\date{\today}% It is always \today, today,

\begin{abstract}

We investigated the complete thermodynamic cycle of aluminium nanoparticles through classical molecular dynamics simulations, spanning a wide size range from 200 atoms to 11000 atoms. The aluminium-aluminium interactions are modelled using a newly developed Bayesian Force Field (BFF) from the FLARE suite, a cutting-edge tool in our field. We discuss the database requirements to include melted nanodroplets to avoid unphysical behaviour at the phase transition.
Our study provides a comprehensive understanding of structural stability up to sizes as large as $3 × 10^5$ atoms. The developed Al-BFF predicts an icosahedral stability range of up to 2000 atoms, approximately 2 nm, followed by a region of stability for decahedra, up to 25000 atoms. Beyond this size, the expected structure favours face-centred cubic (FCC) shapes.
At a fixed heating/cooling rate of 100K/ns, we consistently observe a hysteresis loop, where the melting temperatures are higher than those associated with solidification. The annealing of a liquid droplet further stabilizes icosahedral structures, extending their stability range to 5000 atoms. Using a hierarchical k-means clustering, we find no evidence of surface melting but observe some mild indication of surface freezing. In any event, the liquid droplet's surface shows local structural order at all sizes.

\end{abstract}

\maketitle

\section{\label{sec:level1}Introduction}
The melting and freezing of metallic nanoparticles (MNPs) deviate remarkably from their bulk counterparts because of the interplay between surface effects and solid-solid structural transitions, resulting in a complex free energy surface and different solid-to-solid and solid-to-liquid pathways.  \cite{balettoreview, li2014nanothermodynamics,truhlar2014nanothermodynamics}
\par
Understanding the phase transitions of MNPs is fundamental in designing devices for targeted applications. In particular, Aluminum nanoparticles (AlNPs) have attracted much interest due to their applications in strategic industries, including catalysis\cite{Das202hydroboron} and energy storage\cite{zheng2008hydrogen}. 
\par
To this aim, Molecular Dynamics (MD) simulations are an invaluable tool revealing the atomistic mechanisms of phase transitions at the nanoscale, including surface melting\cite{zeni2021data}. Unfortunately, {\em ab initio} MD is unfeasible due to large timescales and configurational ensembles, while classical MD based on EAM \cite{puri2007effect}, Streitz-Mintmire \cite{puri2007effect}, or second-moment approximation of the tight-binding potentials (TBSMA)\cite{zeni2021data} fails to replicate the melting temperature of the bulk surfaces. A reason for such discrepancy could reside in the fitting over bulk crystal properties, which renders traditional interatomic potentials unable to model the rupture or formation of bonds and the heavy distortion of the lattice structure at high temperatures, with the appearance of many defects. Hence, their applications to AlNPs might provide misleading results. 
\par
Following the seminal works by Parrinello and Behler \cite{Parrinello2007} and Bart\'ok \cite{Bartok2010_GAP}, machine-learnt potentials (MLPs) trained on DFT calculations have emerged as an alternative that bridges quantum and classical methods. MLPs promise to offer accuracy at the level of {\em ab initio} methods at fractions of their computational cost while exhibiting favourable scaling with system size, unlocking simulations of large systems at the timescale of nanoseconds, and have already been successfully used for cMD of Au \cite{jindal2020structural} and Al \cite{chapman2020nanoscale} MNPs (however in the latter a small set of nanoparticles was considered).
\par
Among the various MLPs, Bayesian Force Fields (BFFs) possess many desirable features, most notably the ability to learn efficiently from small datasets and an intrinsic measure of predictive uncertainty, which enables active learning. FLARE ---Fast Learning of Atomistic Rare 
Events\cite{vandermause2022ace,Vandermause2020_sqexp,Owen2023_AuSurface} --- is a state-of-the-art package to perform active learning and training of BFFs and has been successfully applied to a variety of nanosystems, such as Pt and Au nanoparticles and aluminium and gold surfaces \cite{Owen2023_PtNP, zeni2021compact}.
\par
Here, we present a robust BFF for AlNPs trained using FLARE. We compare the performance in reproducing low-index surface energies and bulk melting temperature with respect to available experimental data, providing a comprehensive understanding of the behaviour of AlNPs. 
We then study the energy stability of AlNPs, predicting icosahedra as the most favourable geometry up to 2000 atoms ($\sim$ 2 nm) and of FCC-like motifs behind 25000 atoms. Finally, we simulate the melting/freezing of AlNPs, estimating the presence of a hysteresis loop and the shift of the icosahedral range coming from the annealing of a liquid droplet.
\section{\label{sec:Met}Methodology}
We study the complete thermodynamical cycle, melting/freezing, of Al nanoparticles with various initial geometries between 10$^2$-10$^4$ atoms. We perform iterative Molecular Dynamics (itMD) simulations using LAMMPS.\cite{LAMMPS} itMD is a standard procedure \cite{delgado2021universal,rossi2018thermodynamics}where the system temperature is modified iteratively in finite quantities. After equilibrating the system at an initial temperature, the system temperature is raised or lowered by a fixed amount $\Delta T$ after a specific time interval $t_{itMD}$, long enough to sample the target ensemble at each temperature significantly. The procedure is repeated until the desired final temperature is reached. The heating/cooling rate is simply
$\lambda=\Delta T/t_{itMD}$. 

Here, we consider various starting AlNPs configurations, built as geometrical structures with small random displacements, which are initially equilibrated at 500~K, heated up to 1100 K and then cooled back down to 500 K. A Langevin thermostat regulates the system temperature with a relaxation time $\tau=20$ ps. We fix $\Delta T = 50 \text{K}$, and then the heating rates can be chosen by adjusting $t_{itMD}$. Here we report results obtained using $\lambda=100$ K/ns. 
During each $t_{itMD}$ interval, the system is free to evolve according to Newton's equations of motion, which are integrated using a Velocity-Verlet algorithm\cite{Verlet1967} with a timestep of 2 fs. For sizes up to 2057 atoms, we average our results over three independent simulations.
%\textcolor{magenta}{abbiamo tempo per fare le altre?} 
Interactions between atoms are modelled with a Bayesian Force Field (BFFs), trained on {\em ad hoc} databases, including bulk, surfaces, and small nanoparticles, as summarised in Table \ref{tab:trainset}. 

\subsection{\label{sec:level2}Training and validating Al-BFFs}
We train Al-BFFs using the FLARE package\cite{Vandermause2020_sqexp}, which benefits from integrated active learning capabilities. 
Training is performed on energy, force, and stress.
Active learning trajectories are initiated from an {\em ab initio} calculation on an initial configuration over which a preliminary BFF is trained. The systems evolve according to the preliminary BFF until the predictive uncertainty of at least one local environment exceeds a fixed threshold, at which point DFT calculations are performed, and the potential is updated, then used again to perform MD until the uncertainty threshold is crossed again.
Active learning efficiently samples the system's configurational space by including configurations only when needed, based on the available database.
We use Quantum Espresso (QE) \cite{Giannozzi_2017} as {\em ab initio} reference, selecting the PBEsol functional (QE-PBEsol). Generation of the database and model training are better detailed in Suppl. Info.
\textit{Featurization} of local environments is performed via the Atomic Cluster Expansion (ACE) 3-body B2 descriptors \cite{Drautz2019_ACE} using Chebyshev polynomials of the first kind as the radial basis. 
The kernel function is taken as the Squared Dot Product allowing for mapping onto a quadratic model.\cite{vandermause2022ace}
That choice allows for lossless mapping onto a fast quadratic model whose cost scale linearly with system size:
\begin{equation}
    \epsilon(\textbf{q})=\sigma^2 \textbf{q}^T\xi\textbf{q} ~~~,
\end{equation}
where $\xi$ is a square matrix of the same size as $\textbf{q}$, which is a descriptor vector, while $\sigma$ is an energy parameter which accounts for the variability in atomic energy\cite{vandermause2022ace}.

Dataset 1 ($\mathcal{D}_1$) comprises the DFT calculations obtained by performing active learning on Al bulk, surfaces, and a few small, geometrically-built nanoparticles. It is used to train the potential "Al-BFF 1". A second set of {\em ab initio} calculations on nanoparticles is built to solve the occurrence of non-physical trajectories at high temperatures.
The potential "Al-BFF 2" is trained on the dataset $\mathcal{D}_2$, which contains some configurations from $\mathcal{D}_1$ and configuration of melted nanoparticles/liquid nanodroplets. Al-BFF 2 prevents any of the non-physical behaviour of Al-BFF 1. The content of each dataset is detailed in Table \ref{tab:trainset}, and more details are available in the dedicated section of Suppl. Info..
\par
\begin{table}[h!]
\centering
\begin{tabular}{|c|cccc|}
\hline
Geometry & N\textsubscript{at}   & ${\mathcal D_1}$& ${\mathcal D_2}$ & Test \\ \hline
FCC Bulk & 108 & 36       & 36       & 8    \\ 
BCC Bulk & 128 & 44       & 8        & 9    \\ 
FCC (100)  & 176 & 21       & 21       & 6    \\ 
FCC (110)  & 176 & 32       & 32       & 4    \\ 
FCC (111)  & 176 & 17       & 17       & 6    \\ 
Dh\textsubscript{85}    & 85  & 4        & 4        & -    \\ 
Ih\textsubscript{55}    & 55  & 12       & -        & -    \\ 
Al\textsubscript{100}   & 100 & -        & 3        & -    \\ 
Al\textsubscript{150}   & 150 & -        & 26       & -    \\ \hline
Total    &     & 166      & 147      & 33   \\ \hline
\end{tabular}
\caption{Breakdown of the datasets used to train and test the two BFFs used, indicating the type of structure, the number of atoms N\textsubscript{at} comprising it, how many were included $\mathcal{D}_1$ and $\mathcal{D}_2$, and in the testset}
\label{tab:trainset}
\end{table}

\begin{table}[h!]
    \centering
    \begin{tabular}{|c|cccc|}
    \hline
&  $\sigma$ [eV] & $\lambda_E$ [eV] & $\lambda_f$ [eV/\AA] & $\lambda_\tau$ [eV/\AA\textsuperscript{3}] \\ 
\hline
    Al-BFF 1 &  3.51 & 0.15 & 0.05 & 0.0006 \\
    Al-BFF 2 & 3.51 & 0.15 & 0.06 & 0.0005 \\
     \hline
    \end{tabular}
    \caption{The final values of hyperparameters and trained parameters for both the BFFs used. $\lambda_E$, $\lambda_f$ and $\lambda_\tau$ stand for the energy, force, stress observational noises.}
    \label{tab:params_bff}
\end{table}
Observational noises $\lambda$, which quantify the error tolerance of GPR on the training data, and the energy parameter $\sigma$ are fitted on data, see Table \ref{tab:params_bff}. On the other hand, we keep fixed the ACE hyperparameters, namely the size of the radial and angular bases $n_{max},l_{max}$, and the cutoff radius $r_c$. The values are optimised by training BFF using $\mathcal{D}_1$. We choose the set with the best accuracy bulk FCC predictions, i.e. lattice constant, $a_0$, cohesive energy, $\epsilon_b$, bulk modulus $B$, and a small number of basis functions, see Table \ref{tab:res_bulk}. We choose $n_{max},l_{max}, r_{c}$ as  8, ~3 and 4.5 \AA, respectively, for both Al-BFF 1 and Al-BFF 2.  
\begin{table}[h!]
\centering
$\begin{array}{|c|cccc|} 
\hline
 & \text{Al-BFF 1} & \text{Al-BFF 2} & \text{QE-PBEsol} & \text{Expt.}\\
 \hline
\epsilon_{b} \,{[}\text{eV}{]} & - 3.81 &- 3.82 &- 3.79 &- 3.39 \cite{gaudoin2002} \\
a_{0} \, \text{{[\AA]}} & 4.015 & 4.015 & 4.015 & 4.022 \cite{gaudoin2002}\\
B \, {[}\text{GPa}{]} & 79.8 & 79.9 & 82.7 & 81.3 \cite{gaudoin2002} \\
T_{melt} [\text{K}] & 958 \pm 7 & 906 \pm 1 & - & 933 \cite{AschroftMermin}\\
\gamma_{100} \, {[}\text{J/m\textsuperscript{2}}{]} & 1.06 & 1.10 &  1.09 & -\\
\gamma_{110} \, {[}\text{J/m\textsuperscript{2}}{]}  & 1.15 & 1.19  & 1.15 & -\\
\gamma_{111} \, {[}\text{J/m\textsuperscript{2}}{]}  & 0.90 & 0.94 & 0.96 & 1.16 \cite{boer1988cohesion}\\
\hline
\end{array}$
\caption{Quantities for a selection of properties of FCC Al bulk and low Miller indexes surfaces using different methods or experimentally measured.}
\label{tab:res_bulk}
\end{table}
Table \ref{tab:res_bulk} lists some static properties at 0 K, such as elastic constants, bulk moduli, and surface energies $\gamma$ for low-index facets, as well as the predicted melting temperatures $T_{melt}$, for BFFs, QE-PBEsol, and experimental references.
\begin{table}[h!]
    \centering
    \begin{tabular}{|c|c|c|c|}
    \hline 
    & Al-BFF 1  & Al-BFF 2 & QE-PBEsol\\
    \hline Forces $\rho$ &  0.996   & 0.996 & - \\
    \hline Forces MAE [eV/\AA] & 0.024 & 0.026 & - \\
    \hline Forces MAV [eV/\AA]&0.2504  & 0.2480 & 0.2549 \\
    \hline Energies $\rho$ & 0.998 & 0.997 & - \\
    \hline Energies MAE [eV]& 0.003 & 0.005  & - \\
    \hline Energies MAV  [eV]&  3.677 & 3.681  & 3.676 \\
    \hline \end{tabular}
    \caption{Pearson's $\rho$ and MAE of predicted and computed quantities of the training set, for two different Al-BFFs versus our reference {\em ab initio} calculations, as in QE and PBEsol.}
    \label{tab:gp_mae}
\end{table}
To further assess the goodness of the resulting potentials, we compute the mean absolute error (MAE) they incur in predicting total energies and the component of atomic forces on some configurations that were randomly picked from the initial active learning trajectory and excluded from the training set, as well as Pearson's correlation coefficients $\rho$ between predicted and reference data and the median absolute values (MAV), as listed in Table \ref{tab:gp_mae}. 

The melting temperature of bulk Al is evaluated, for both Al-BFFs and Mishin-EAM\cite{mishin1999eam} with the interface velocity method \cite{morris1994melting}, where NPT simulations of a system comprising both the solid and liquid phase are performed at different temperatures. The melting temperature is taken as the one where neither melting nor crystallization occurs. In that way, the velocity of the interface between the two phases is zero. 
Al-BFF2 predicts a bulk melting temperature of 906 $\pm$ 1 K which slightly underestimates the experimental value of 933 K, but in any event is closer than the Physically Informed NN's (PINN) prediction of 975 K\cite{Mishin_2020NN}, and especially of the EAM at 1041 K, indicating that its prediction of melting properties is more accurate.
Performance of BFF and QE-PBEsol are benchmarked simulating 108 Al atoms in the FCC bulk. The former was $2.2 \cdot 10^4$ atom step s\textsuperscript{-1} while the latter was $4 \cdot 10^{-2}$ atom step s\textsuperscript{-1}, resulting in a speed-up of six orders of magnitude.

\subsection{Characterisation of phase transition in AlNPs}
The thermodynamic data and the trajectories obtained from itMD are analysed using standard energetic and structural quantities to understand the melting/freezing transition. Structural quantities are calculated from an adapted version of the \texttt{Sapphire}\cite{jones2023structural}.
 
The excess energy of a nanoparticle is the difference between its total energy and that of a bulk portion of the same nuclearity, $N$, weighted by the rough estimate of surface atoms according to the Spherical Cluster Approximation (SCA) :
    \begin{equation}
    \Delta E = \frac{E(N)-N\epsilon_b}{N^{2/3}} ~~~,
\end{equation}
where $\epsilon_b$ is the bulk cohesive energy and $E(N)$ the total energy of the AlNPs.
$\Delta E$ is defined positive, and represents the energy cost of carving the nanoparticle out of the bulk material. The smaller $\Delta E$ is, the more stable the nanoparticle is.\cite{balettoreview}
The specific heat per atom is computed using the fluctuation-dissipation theorem,
\begin{equation}
    c_{p=0}(T) = \frac{\langle E(N)^2 \rangle - \langle E(N) \rangle^2}{N k_b T^2}~~~,
\end{equation}
where $T$ the nominal temperature and $k_b$ the Boltzman constant.
The pair distance distribution function, PDDF $p^{(2)} (d)$, is defined as the probability that a certain atomic pair is at a distance $d$:
\begin{equation}
    p^{(2)} (d)= \frac{2}{N(N-1)} \sum_{i>j} \delta(|\textbf{r}_i - \textbf{r}_j|-d) ~~~,
    \label{eq:pddf}
\end{equation}
and provides information on the existence of a geometrical order.\cite{delgado2021universal}

While the first peak of the $p^{(2)}(d)$ corresponds to the position of the nearest neighbor, the position of the second peak labels whether the nanoparticle has a geometrical order or not. If the second peak falls at the lattice distance, the nanoparticle has a geometrical order; alternatively, it is amorphous or melted.\cite{delgado2021universal}

Another useful distribution informative about the geometry of the nanoparticle is the radial distribution function (RDF), $p(r)$. $p(r)$ is the probability of finding an atom at a distance $r$ from the centre of mass (COM, positioned at $\textbf{r}_0$), 
\begin{equation}
    p(r)=\frac{1}{N}\sum_i \delta(|\textbf{r}_i-\textbf{r}_0|-r) ~~~.
\end{equation}
During the melting, one could observe a radial expansion, inflation, of the nanoparticle, in agreement with experimental observation.\cite{delgado2021universal}
As a good estimate of the NP radius we take the gyration radius
\begin{equation}
    r_{gyr}= \sqrt{\frac{1}{N} \sum_i |\textbf{r}_i - \textbf{r}_0|^2} ~~~,
\end{equation} 
with obvious meaning of the symbols. A rough estimate of the radius of a nanoparticles is given by the Spherical Cluster Approximation, where it is taken as a sphere where every atom occupies the same volume as in the FCC bulk at 0 K, then $r_{SCA}=(N)^{1/3} a_0/\sqrt{8}$~ which in our case $\sim 1.42 (N)^{1/3}$. To compare radii of nanoparticles of different size, we the use an adimensionalized ratio $\rho= r_{gyr}/r_{SCA}$.
\\
%i really hope radice(2)/4 = 1/radice(8)
The Kullback-Leibler (KL) divergence is a distance in the space of distributions. The divergence between different PDDFs at different temperatures quantifies the structural order:
\begin{equation}
    D_{KL}(p_1^{(2)}|p_2^{(2)})= \int p_1^{(2)}(x) \log{\frac{p_1^{(2)}(x)}{p_2^{(2)}(x)}} dx ~~~.
\end{equation}
It has been shown that the KL divergence between a reference solid PDDF and the one of nanoparticles at higher temperatures presents a quasi-first order transition at the phase-change temperature\cite{delgado2021universal}. 
Therefore, we define the melting and the freezing temperature, $T_m$ and $T_f$ as the temperatures showing the nearest larger and smaller $D_{KL}$ to the discontinuity.

Common Neighbour Analysis (CNA), introduced by Honeycutt \textit{et al.}, labels each pair of atoms with a signature based on the connectivity between their common neighbours \cite{Honeycutt1987CNA, baletto2019jpcm}.
While only a small fraction of CNA signatures are needed to identify solid NP-shapes, namely (555), (421) and (422), for the fivefold symmetry axis, FCC-bulk environment and stacking fault planes, respectively, unsupervised learning helps classify different states occurring during the melting and freezing cycle. To calculate the CNA signature we use a fixed cutoff distance of $0.8 a_0$.  We adopt the hierarchical k-means clustering to isolate classes of local atomic environments based on ACE B2 descriptors, as described in Zeni \textit{et al.}\cite{zeni2021data} and Jones \textit{et al.}\cite{jones2023structural} and implemented in the \texttt{Raffy} package \cite{jones2023structural}.
\begin{figure}[ht!]
    \centering
\includegraphics[width=0.5\textwidth,keepaspectratio]{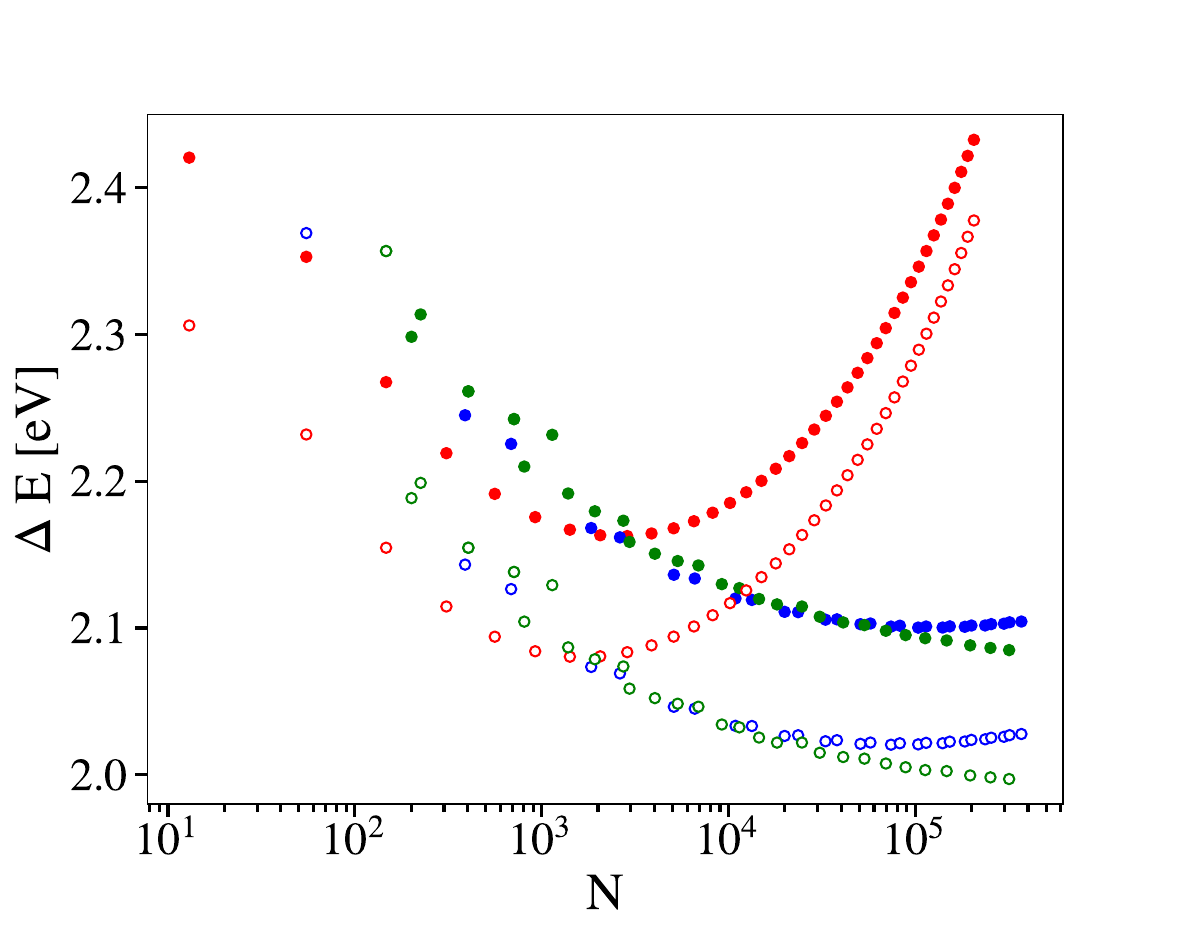}
    \caption{Excess energy in eV as a function of the nanoparticle size using Al-BFF 1 (empty circle) and Al-BFF 2 (full circle). Red points are icosahedra (Ih), blue points are Mark Decahedra (MDh) while green points are truncated octahedra.}
    \label{fig:delta_en}
\end{figure}
\section{Results}
We calculate the excess energy for the three main structural families, namely icosahedra (Ih), decahedra (Dh) and FCC polyhedra - such as octehadra, Oh, truncated octahedra, TOh, and cuboctahedra, COh- with both our BFFs as shown in Fig. \ref{fig:delta_en}. Structures are relaxed until all force components are smaller than $10^{-5}$ eV/\AA.

The best structures for Dh turned out to be Mark Decahedra with $m=n$ and $p=n/2$ in agreement with other reports \cite{baletto2002crossover}. the best FCC polyhedra respect the Wulff construction with $\gamma_{100}/\gamma_{111}=1.17 \sim d_{100}/d_{111}~$, where $d_{100}$ and $d_{111}$ are the distance of (100) and (111) facets from the center of mass, respectively. 
%Wulff polyhedra maximise the extent of FCC $(111)$ facets, whose surface energy is the lowest among low Miller index cuts, with . 
The energetics of small clusters is dominated by their surface energy with the icosahedral packing is favoured because of its low surface-to-volume ratio.
The larger the number of atoms, the more lattice strain is prevalent and bulk-like structures such as twinned planes (present in Dh) and FCC cuts become energetically advantageous \cite{balettoreview}. 
There is a disagreement between Al-BFF 1 and Al-BFF 2, with $\Delta E$ being shifted to higher values in Al-BFF 2 than Al-BFF 1. The locations of the energy crossings change too, as shown in Table \ref{tab:size_cross}, with a smaller size-window for Dh predicted with Al-BFF 1 level. Al-BFF 2 tends to stabilise at larger sizes in five-fold symmetry with Dh and FCC polyhedra in close competition up to 10$^5$ atoms.
\begin{table}[h!]
    \centering
    \begin{tabular}{|c|c|c|}
    \hline
        BFF &  Ih $\rightarrow$  Dh/FCC & Dh/FCC $\rightarrow$ FCC \\
        \hline
        Al-BFF 1 & 1500 & 10000 \\
        \hline
        Al-BFF 2 & 2100 & 25000 \\
        \hline
    \end{tabular}
    \caption{Sizes of energy-crossing points between geometrical motifs for Al-BFF 1 and Al-BFF 2.}
    \label{tab:size_cross}
\end{table}
\subsection{Thermodynamical cycle}
\begin{figure*}[ht!]
    \centering
    \includegraphics[width=1\textwidth]{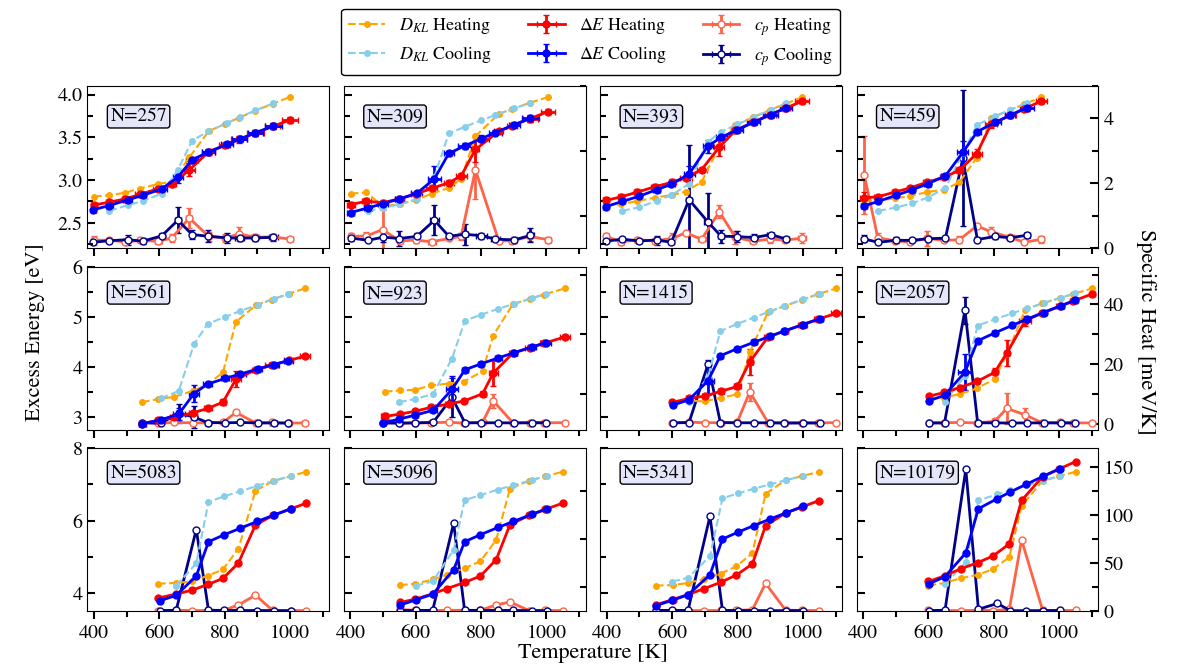}
    \caption{Caloric plots in heating (red) and freezing (blue) for nanoparticles of different sizes using Al-BFF 2 using the itMD procedure with a rate of 100 K/ns. Data averaged over three independent simulations. KL divergence is expressed in arbitrary units.}
    \label{fig:calorics}
\end{figure*}

Fig. \ref{fig:calorics} shows the excess energy, the specific heat per atom, and the PDDF-KL divergence as a function of the temperature. The  PDDF-KL divergence is calculated with respect to the PDDF at the end of the annealed process.

For nanosystems, the melting occurs on a range of temperatures, leading to finite peaks in the specific heat, which become sharper as the size of the system increases, approaching their bulk limit. The transition is sharper in icosahedra than in decahedra also at similar sizes, in agreement with the predicted energy stability.
By taking a closer look at the caloric plots, we can clearly notice the size-dependent effects that take place:
\begin{enumerate}
    \item The peak in specific heat is sharper for larger AlNPs; 
    \item The melting and freezing temperatures shift upward with increasing size;
    \item The heating and melting curves show larger hysteresis at larger size. 
\end{enumerate}

The hysteresis in the caloric curves, in other words, the difference between $T_m$ and $T_f$, probably has a kinetic origin due to the fast heating rates used in the simulations.\cite{rossi2018thermodynamics} On the other hand, in experiments, the heating is supposed to be quasi-static. 
The Al$_{257}$ hysteresis is almost negligible, while in Al$_{10179}$ we notice a discrepancy of around 200 K between melting and freezing.
The annealed structures are generally more compact and stable than the initial configurations. We note that the latter are not the global minimum, which explains why $c_p$ has a sharper peak during cooling than during melting.

We note that Al$_{309}$, Al$_{393}$, and Al$_{459}$ show a smaller $c_p$ peak during heating at lower temperatures that also correspond to small jumps in the prevalence of some CNA signatures, indicating the presence of geometrical rearrangements reasonably due to the initial deformation.
\begin{figure}[h!]
    \centering
    \includegraphics[width=1\linewidth]{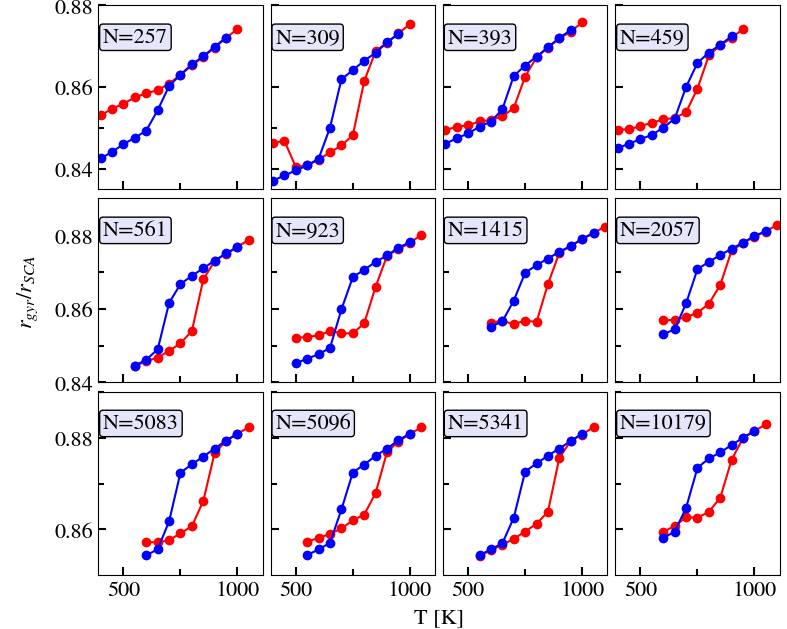}
    \caption{Normalised gyration radius, $\rho$  for the selected sizes as a function of temperature. Red and blue plots refer respectively to heating and cooling half-cycles.}
    \label{fig:plot_radi}
\end{figure}
The hysteresis is evident in any geometrical descriptor plotted versus temperature, see for example the normalised gyration radius $\rho$ in Fig. \ref{fig:plot_radi}. While $\rho$ behaves as expected during cooling, with a linear dependency on temperature at usually a steeper slope than the liquid form, we note an anomalous behaviour in melting. This trend can be explained by the occurrence of geometrical rearrangements of the initially rattled configurations. This is especially the case at the beginning of the Al$_{309}$ heating, where a rapid contraction of the nanoparticle occur, but remains roughly constant in the 900-2000 range. Indeed, the following CNA analysis, see later, show that  Al$_{309}$ undergoes to geometrical reordering in five-fold axis, rather than expanding the intershell distances.

To take into account the hysteresis during the full thermodynamical cycle, we contrast the experimental melting temperatures of Al-NPs with the average value $(T_m+T_f)/2$.  Size-dependent properties, $A(N)$, of metallic NPs are expected to follow a Gibbs-Thomson scaling law \cite{guisbiers2010size}:
\begin{equation}
    A(N)= A \left(1-\frac{C}{N^{1/3}}\right)^{s} ~~~,
\label{eq:scaling}
\end{equation}
where $N$ is the NP size and $A_{\infty}$ is the corresponding bulk value. The constant $C$ and the exponent $s$ depends on the considered property, and is 1 for melting/freezing. $C$ relates to the surface effects of a certain material, and it may show a dependency on the NP-shape.\cite{guisbiers2010size, truhlar2014nanothermodynamics}
Fitting of Eq. \ref{eq:scaling} yields to a value of 1.58 for $C$, close to that obtained by fitting experimental data\cite{lai1998melting} which amounts to 1.76. A plot of the phase transition temperatures as a function of nanoparticle size is provided in Suppl. Info./Appendix.
The good agreement we have with the experimental estimate for $C$ corroborates the accuracy of our Al-BFF 2 potential and its reliability in predicting melting/freezing behavior.

\begin{figure}[ht!]
    \centering
    \includegraphics[width=1\linewidth, height=0.4\textheight]{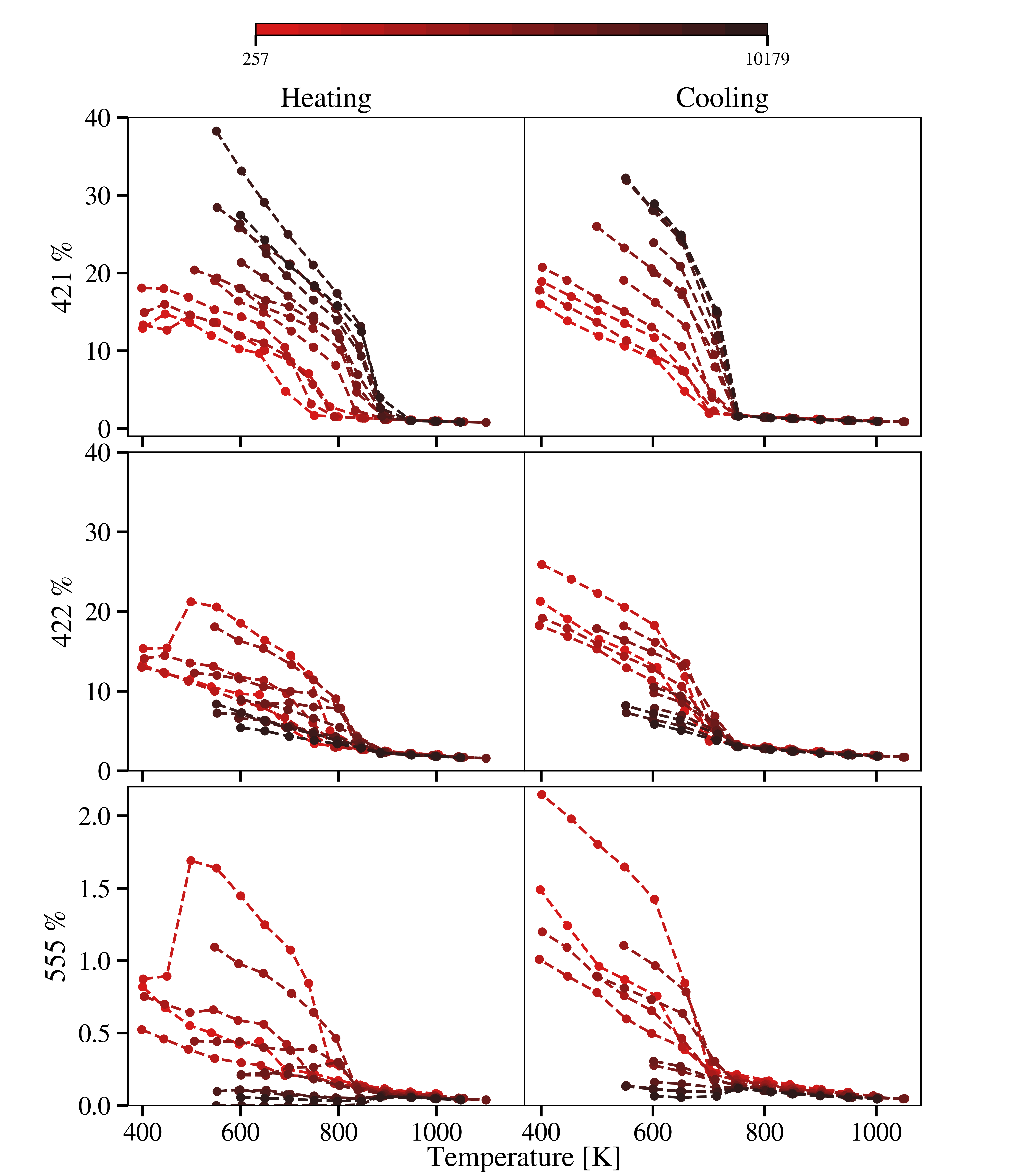}
    \caption{Occurrence of the (421), (422) and (555) CNA signatures as a function of temperature. The left column reports values for the heating process, the right column for the cooling one. The line-color refers to the Al-NP size accordingly to the legend.}
    \label{fig:sign}
\end{figure}

\begin{figure}[h!]
    \centering
    \includegraphics[width=0.85\linewidth]{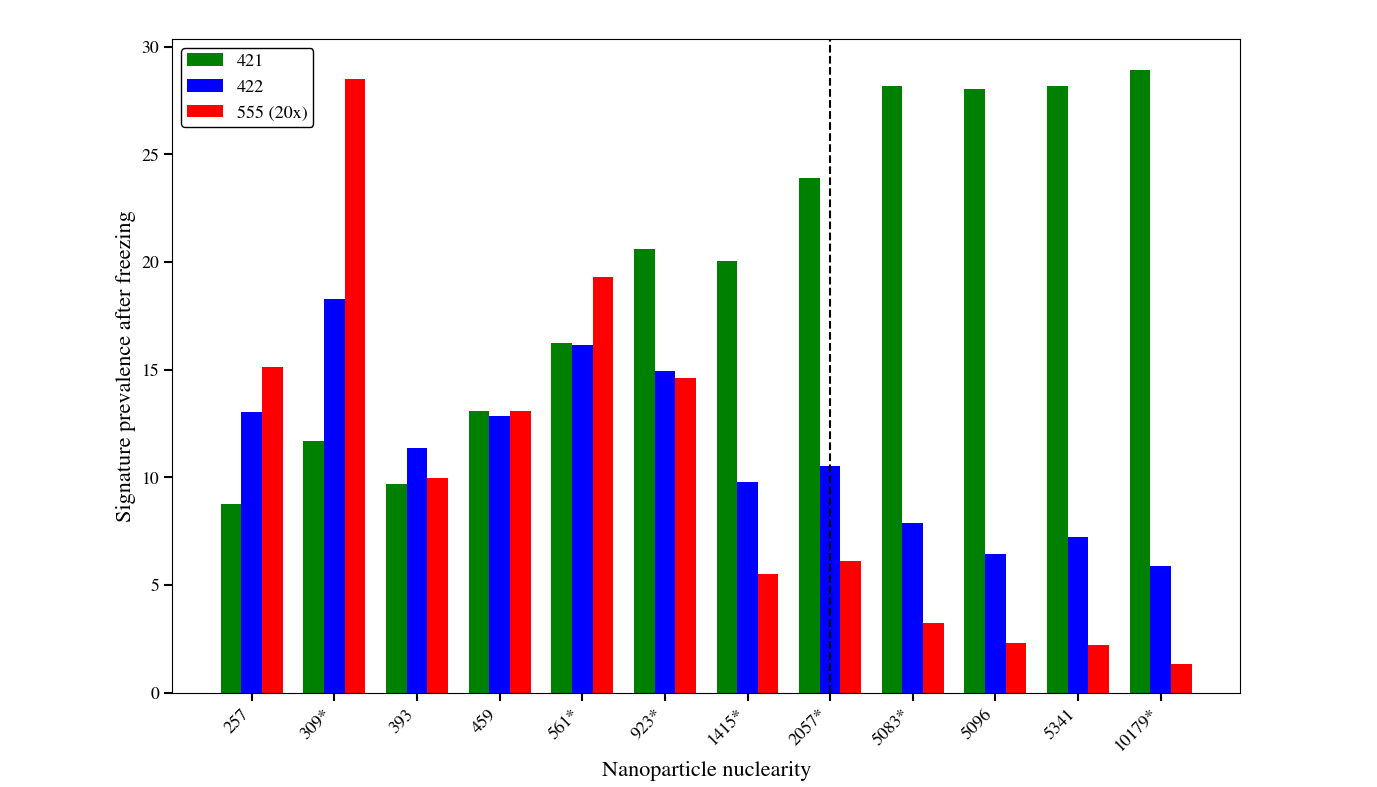}
    \caption{CNA signatures at 600 K, after freezing averaged over available trajectories. Green bars refer to (421) \%, blue bars to (422) and red ones to (555). The latter occurrence is multiply  by 20 times for being visible on the graph. Sizes corresponding to the geometrical closure of icosahedra are indicated by an asterisk. The vertical dashed line indicates the predicted size at which we expect the Ih$\rightarrow$Dh from the energy profile, see Table \ref{tab:size_cross}.}
    \label{fig:sign2}
\end{figure}

To classify the structure family during the thermodynamical cycle, we propose a CNA analysis.\cite{balettoreview} Fig. \ref{fig:sign} reports 422, 421, and 555 CNA-signature occurrence against temperature, while Fig. \ref{fig:sign2} quantifies the occurrence of CNA signatures at 600 K. We select that temperature because it is the first value where all the Al-NPs are solid independently of their size. 

We use Fig. \ref{fig:sign2} for a fast classification of the family type. We split only between icosahedra, decahedra, and FCC-like, as in Ref.\onlinecite{rossi2018thermodynamics}.
We note that for liquid droplets, all the three CNA signatures reach the same value independently of the NP-size. As expected, the (421) \% of solid AlNP increases considerably with size, while (555) and (422) \% decrease with the NP nuclearity. The (555) approaches zero for the two largest sizes considered. The jump at low temperature for N=257 is due to the structural reordering towards Ih discussed before.

\par
The 600~K- distribution of CNA signatures confirms the downward trend of (555) and (422)\%, with the (421)\% becoming dominant, as expected.\cite{rossi2018thermodynamics}
Up to 923 Al nanodroplets solidify into Ih, in agreement with the energetic profile. 
Increasing the AlNP size, the (555) \% dwindles to zero, but it becomes small only after 5100 atoms. 
We note that the occurrence of (421) is greater than the (422) for AlNPs with more than 1000 atoms, but the latter remains within the 5-10\% of the atomic pairs with such a local environment. This feature translates into the formation of defective shapes with several dislocation planes, often not parallel. Furthermore, a non-zero (555) signature highlights the formation of fivefold shapes, as Ih and Dh.\cite{balettoreview} 
Indeed, the (421) \% is not high enough to suggest a structural change after 5100 atoms. Al$_{5083}$, Al$_{5096}$, and Al$_{5341}$ solidify as defected-Ih, with an incomplete fivefold axis, the reason why the (555)\% is smaller. In particular, at 5341 atoms, we would expect a Wulff polyhedra, but we obtain an Ih with some vacancy around the fivefold vertex. Al${10179}$ solidifies into a defected Dh with the fivefold axis not centred.

As it could be of interest in the field of 3D printing\cite{li20123d}, we use a hierarchical k-means clustering to highlight where the melting/solidification starts. We use a similar approach to check the surface melting in Au nanoparticles. \cite{zeni2021compact}
\begin{figure}[ht!]
    \centering
    \includegraphics[width=1.05\linewidth]{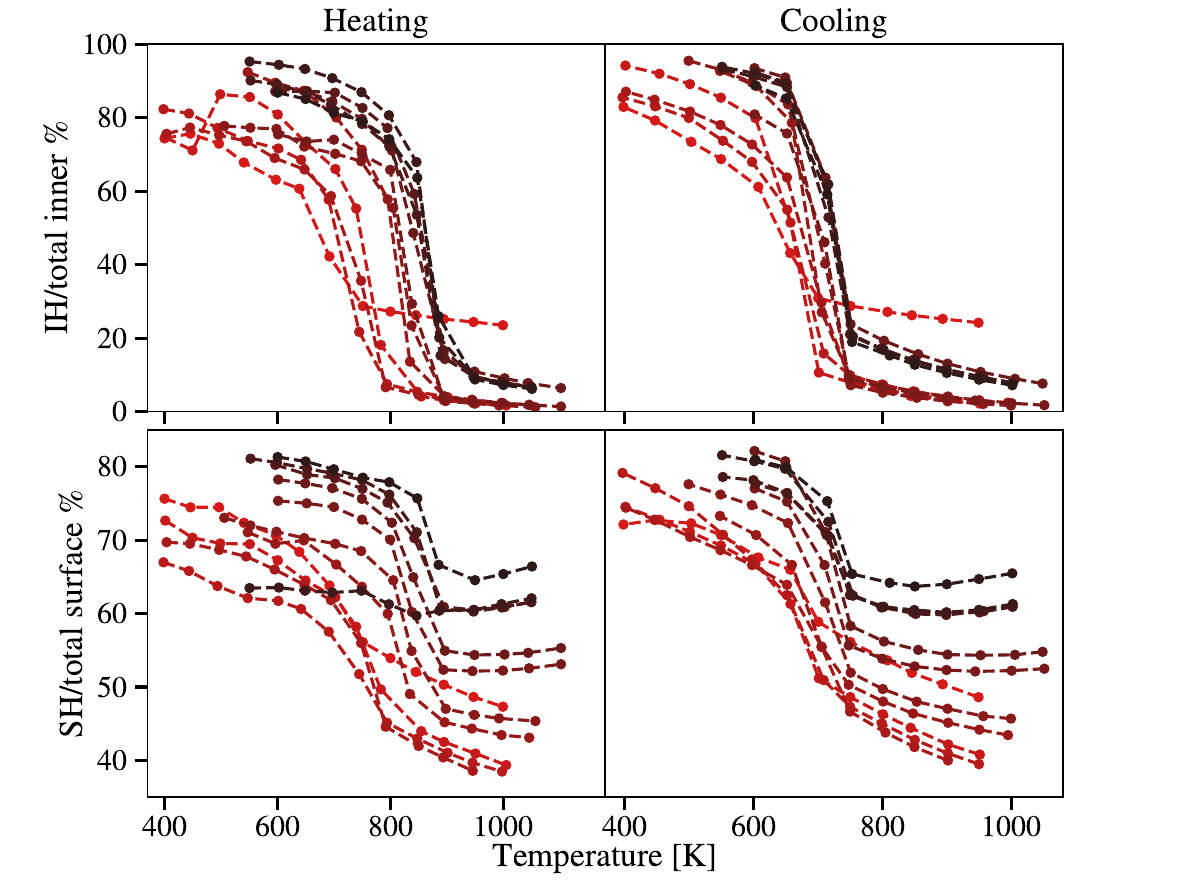}
    \caption{Occurrence of ordered environments in the inner and at the surface of Al-NPs as detected using K-means clustering, as a function of temperature and number of atoms. Left column stands for heating while right column during cooling. Colour coding as in Fig. \ref{fig:sign}.}
    \label{fig:raffy_labels}
\end{figure}
Hierarchical k-means clustering is performed using $12663$ local environments as training set, randomly chosen from the obtained cMD trajectories. The clustering follows the \texttt{RAFFY} library, see Jones \cite{jones2023structural}, and the local environments are defined by 40 B2 functions similar to our previous work, Zeni et al. \cite{zeni2021compact}. We perform two tiers of clustering. First, we distinguish between inner (I) and surface (S) atoms, while T+the second tier differentiates between environments with or without local order, as can be inferred by the difference in their average coordination number (ACN). This defines 4 different types of environment: inner and ordered (IH) with an ACN of 12, inner and disordered (IL) with ACN 11, surface and ordered (SH) with ACN 8, surface and disordered (SL) with ACN $\leq$7. The plots containing the occurrence of these labels at every snapshot are included in the Suppl. Info.

The percentage of inner and surface atoms that show local order in Al-NPs of different sizes is plotted in Fig. \ref{fig:raffy_labels} as a function of temperature during the heating and cooling half-cycles. As expected, the occurrence of ordered or highly coordinated environments drops after 750~K.
The percentage of ordered inner (IH) atoms drops/peaks at the transition temperature during heating/freezing, respectively. Changes in the core are steeper than at the surface, with the latter spread over a range of temperatures and never approaching zero. Interestingly, during freezing, the IH percentage rises below 750~K independently of the NP-size. On the other hand, it spreads between 700-850~K during heating. Furthermore, we note an increment in the IH percentage sharper than that of SH, indicating that the surface organises first than the inner part. We do not observe any surface melting. In fact, the change in the percentage of inner and surface solid environments occurs at the same temperature, in agreement with other studies \cite{tran2016surface,levitas2011size}.

\section{Conclusion}
We developed a Bayesian force field (BFF), employing the FLARE suite to investigate aluminium nanoparticles' thermodynamical cycle using classical molecular dynamics simulations.
We demonstrate the need to include melted nanoparticles in the dataset to improve the prediction of AlNPs' trajectories at various temperatures. Our simulations confirm that a dataset containing sizes as small as 85, 100, and 150 atoms is sufficient to explore a 10 to $10^5$-atoms range for energetic and between 200-12000 atoms for the thermodynamical cycle. 
From a standard energy analysis, icosahedral shapes are energetically favourable with respect to decahedra and FCC-polyhedra up to 2100 atoms without any formal global minimisation. Among FCC-polyhedra, Wulff truncated octahedra are the most favourable and predominant at sizes above 25000 atoms. However, a hysteresis loop is always at a fixed rate of 100 K/ns during the heating and cooling cycle. The loop enlarges, increasing the nanoparticle size. 
Our results align with the knowledge granted on AlNPs but improve the melting temperature prediction for bulk, surface, and Al-NPs, getting them in close agreement with the experimental data. Our results agree very well with the Gibbs-Thomson fitting of a phase-transition temperature similar to that of the experimental data.
We also elucidate significant differences between the freezing and melting mechanisms. Heating and cooling half-cycles are not reversible, and the latter steeps at lower temperatures almost simultaneously in the inner and the surface of the Al-NP. During heating, using the terminology offered by Truhlar et al.\cite{truhlar2014nanothermodynamics}, the nano-slush state is at least up to 3.4 nm, as predicted by various geometric and energetic descriptors. 

From a structural point of view, the proposed CNA analysis during the complete thermodynamical cycle supports the stability of icosahedral shapes on a broader size range of up to 5000-6000 atoms. Only Al$_{10179}$ adopts a non-icosahedral shape after freezing.

This work will promote the use of Bayesian force fields to investigate dynamical and thermodynamical properties at the nanoscale, inspiring further research and potential applications in materials science and nanotechnology.

\begin{acknowledgments}
We re grateful for the financial support from the European Commission, under the contract EIC Pathfinder CHIRALFORCE \#101046961. 
DA thanks the University of Milan and ICCOM-CNR for financially supporting his PhD studentship (D.M. n. 630/2024 PNRR).
Computational resources are provided by the INDACO Platform, an HPC project at the University of Milan.

\end{acknowledgments}

\section*{Author Contributions}
FB had the original idea, DA performed the simulations. Both authors contribute in the data analysis and writing.
\section*{Conflicts of interest}
There are no conflicts to declare.

\section*{Data Availability Statement}

\begin{center}
\renewcommand\arraystretch{1.2}
\begin{tabular}{| >{\raggedright\arraybackslash}p{0.3\linewidth} | >{\raggedright\arraybackslash}p{0.65\linewidth} |}
\hline
\textbf{AVAILABILITY OF DATA} & \textbf{STATEMENT OF DATA AVAILABILITY}\\  
\hline
All our data, including the parametrization of the two Al-BFF and the cMD trajectories are public.
&
The parameters of Al-BFF 1 and 2 and the database of QE-PBEsol calculations used to train and validate them. The classical MD trajectory used within the article [and its supplementary material] are openly available through the public repository provided by our institution, UNIMI Dataverse at DOI 10.13130/RD\_UNIMI/UTVVS8 under a CC-BY SA 4.0 licence.
\\
\hline
\end{tabular}
\end{center}

%\nocite{*}
\bibliography{article}
% Produces the bibliography via BibTeX.

\appendix

\section*{Supplementary Information}
\subsection{DFT calculations for Al \label{ss:dft}}
We use Quantum Espresso \cite{Giannozzi_2017} for the ab initio calculations, electing PBEsol as our exchange-correlation functional \cite{Perdew2008_PBEsol} in the Ultrasoft Pseudopotential form as computed by Dal Corso \cite{dal2014pseudopotentials}. The cutoffs on the wavefunctions and charges are, respectively, $400$ eV and $2400$ eV. A Monkhorst-Pack grid \cite{monkhorst1976special} was used for reciprocal space integration with the condition that for each periodic dimension $n_\alpha L_\alpha > 80$ \AA. A Marzari-Vanderbilt \cite{Marzari1999} smearing of the occupation functions is introduced with a smearing width of $0.1$ eV.

\subsection{Active learning}
Active learning trajectories of bulk and surface structures were performed at 700 K, while clusters were simulated at 100 K. A lower threshold for adding atoms to the sparse set, $\sigma_{add}$ was set, so that a reasonable number of environments after every DFT call. $\sigma_{DFT}$ and $\sigma_{add}$ were chosen respectively 0.001 and 0.0005 in units of $\sigma$.In FLARE, while the noises and $\sigma$ are optimised by maximising the log-likelihood, the hyperparameters $n_{max},l_{max}$ and $r_{c}$ are fixed by the user. Parameters for initial active learning trajectories are chosen completely arbitrarily.  For the ACE descriptors, $n_{max}=9, l_{max}=3, r_{cut}=5$ \AA~are chosen. The uncertainty threshold for calling DFT , $\sigma_{DFT}$ is 0.001 in units of $\sigma$, which is initialized at 3.5 eV. \par

\subsection{Bayesian force field: training and validation \label{ss:validating}}
Our assumption is the locality of the interatomic interaction, enabling us to define the total energy as the sum of atomic energies depending only on the coordinates of other atoms falling within a specific cutoff $r_c$. The collection of distances to these atoms is called the "local environment" of the $i$-th atom, $\boldsymbol{\rho}_i$, and the total energy is written as:
\begin{equation}
    U = \sum_i^{N} \epsilon(\boldsymbol{\rho}_i) ~~~,
\end{equation}
where $N$ is the number of atoms in the system. We model $\epsilon(\boldsymbol{\rho})$ as a Bayesian Force Field (BFF), where the local energies are found as Gaussian Processes (GP) of the local environments:
\begin{equation}
    \epsilon(\boldsymbol{\rho}) \sim \mathcal{GP}(\mu(\boldsymbol{\rho}),k(\boldsymbol{\rho},\boldsymbol{\rho}')) ~~~.
\end{equation}
A GP is the generalisation of Multivariate Gaussian Distributions to a number of infinite variables. GP is a distribution of functions, whose mean is $\mu(\boldsymbol{\rho})$ and covariance, or kernel, function is $k(\boldsymbol{\rho},\boldsymbol{\rho}')$ \cite{Rasmussen_GP_ML}. 

Descriptors $\textbf{q}(\boldsymbol{\rho})$ are functions of pair-distances within a local environment. The local environment contains the required physical information and can be used in place of atomic coordinates to improve learning. 
Gaussian Progress Regression (GPR) is the training of a GP to replicate a target function, such as $\epsilon(\textbf{q})$ by conditioning it on a finite set of known input-output couples in a Bayesian framework.

Given a dataset $\mathcal{D}_{\epsilon}=\{\textbf{q}_{d},\epsilon(\textbf{q}_{d})\}$ containing known energies for some descriptor values, our prediction $\epsilon_{*}$ for the energy of a test environment $\textbf{q}_{*}$ , conditioned on the observations, will be:

\begin{equation}
\epsilon_{*}|\textbf{q}_{*},\mathcal{D}_{\epsilon}\sim\mathcal{N}(\bar{\epsilon}(\textbf{q}_{*}),\text{VAR}(\textbf{q}_{*}))\label{eq:cond_noise}
\end{equation}
where:
\begin{flalign}
\bar{\epsilon}(\textbf{q}_{*})&=\textbf{k}^{T}C\boldsymbol{\epsilon} ~~~, \\
\text{VAR}(\textbf{q}_{*})&=k(\textbf{q}_{*},\textbf{q}_{*})+\lambda^{2}-\textbf{k}^{T}C^{-1}\textbf{k} ~~~, \\
\textbf{k}^{T}&=(k(\textbf{q}_{*},\textbf{q}_{1}),...,k(\textbf{q}_{*},\textbf{q}_{D_{\epsilon}})) ~~~, \\
\boldsymbol{\epsilon}_i &= \epsilon({\boldsymbol{q}_i}) ~~~, \\
C_{pq}&=k(\textbf{q}_{p},\textbf{q}_{q})+\lambda^{2}\delta_{pq} ~~~,
\label{eq:predict_simple}
\end{flalign}
being $\lambda$ a noise parameter. We can also define the Gram matrix $K_{pq}=k(\textbf{q}_p, \textbf{q}_q)$.
The $\bar{\epsilon}(\textbf{q}_{*})$ is the estimate of
the target function that minimises the squared error loss function $\mathcal{L}=(\bar{\epsilon}(\boldsymbol{q})-\epsilon)^{2}$.
The Representer Theorem \cite{kimeldorf1970correspondence} states that Eq. \ref{eq:cond_noise} is equivalent to :
\begin{flalign}    
\bar{\epsilon}(\textbf{q}_*)&= \sum_i \alpha_i k(\textbf{q}_i,\textbf{q}_*) = \boldsymbol{\alpha}\cdot \textbf{k}~~~, \\
\label{eq:predict_weights}
    \boldsymbol{\alpha}&=(K+\Lambda )^{-1} \boldsymbol{\epsilon} ~~~,
\end{flalign}
where $\Lambda = \lambda^2 I$. It is important to stress that every linear function of a GP is itself a GP \cite{Rasmussen_GP_ML}.
Because of the non-locality of Schr\"odinger equations, atomic energies are not well defined in DFT calculations but are themselves obtained as GPs of total energies\cite{GlielmoZeni_2020} imposing $E = \sum_i \epsilon_i$. Forces and energies are well defined in {\em ab initio} calculations and can be included as training data. We can define kernels relating different properties linked by linear operators by $k^{AB}(q_A,q_B')=Lk^{AA}(q_A,q_A')$, where $A$ and $B$ are two different properties, if $B = LA$ where $L$ is a linear operator \cite{deringer2021gaussian}. $K$ and $\Lambda$ become:
\begin{equation}
K=\left[\begin{array}{ccc}
K^{\epsilon\epsilon} & K^{\epsilon\textbf{f}} & K^{\epsilon\tau}\\
K^{\epsilon\textbf{f}^{T}} & K^{\textbf{f}\textbf{f}} & K^{\textbf{f}\tau}\\
K^{\epsilon\tau^{T}} & K^{\textbf{f}^{T}\tau^{T}} & K^{\tau\tau}
\end{array}\right]~~,~~ 
\Lambda= \left[\begin{array}{ccc}
\Lambda_\epsilon^2 & 0 & 0\\
0 & \Lambda_\textbf{f}^2  &0\\
0 &  0 & \Lambda_{\tau}^2
\end{array}\right]~~~,
\end{equation}

Furthermore, these can be plugged in Eq. \ref{eq:predict_weights} by using the column stacking of all the available labels, $\textbf{y}$, instead of $\boldsymbol{\epsilon}$:
\begin{equation}    
\textbf{y}=[\epsilon(\textbf{q}_1),\dots,\epsilon(\textbf{q}_{D_{\epsilon}}), \textbf{f}(\textbf{q}_1),\dots,\textbf{f}(\textbf{q}_{D_f}),\tau(Q_1),\dots,\tau(Q_{D_\tau})] ~~~,
\label{eq:vec_labels}
\end{equation}
then the summation in Eq. \ref{eq:predict_weights} has to perform over different kernels. This leads to new formulations for the predictions in Eq. \ref{eq:cond_noise}, reported here from Glielmo \textit{et al.}\cite{GilelmoDeVita2017}:
\begin{flalign}
\bar{\epsilon}(\textbf{q}_*)&=\sum_{ij} k^{\epsilon t_i}(\textbf{q}_*,\textbf{q}_i) C_{ij}^{-1} y_j ~~~, \label{eq:multiprop_mean}\\
\text{VAR}(\textbf{q}_*)&=k(\textbf{q}_*,\textbf{q}_*) - \sum_{ij} k^{\epsilon t_i} (\textbf{q}_*,\textbf{q}_i) C_{ij}^{-1} k^{t_j \epsilon}(\textbf{q}_j, \textbf{q}_*) ~~~, \label{eq:multiprop_var}
\end{flalign}
where $i$ and $j$ run over all the descriptors (atomic and global) in the training set, $t \in \{\epsilon,\textbf{f},\tau\}$ represent the type of label and $C=(K+\Lambda)$. For clarity, we assume that all quantities are of adequate order (scalar, vectors, 2D-tensors,$\dots$)

Predictions performed with Eqs. \ref{eq:multiprop_mean} and \ref{eq:multiprop_var} present scaling respectively $O(N^2)$ and $O(N^3)$ which makes them unusable on large datasets. To avoid such issue, approximated models called Sparse Gaussian Processes can be used, which rely on a subset of cardinality $M<N$, called an "active set"  \cite{QuinoneroRasmussen2005_SGP,Rasmussen_GP_ML}. Here, we 
use the Deterministic Training Conditional (DTC) approximation to speed up the predictive uncertainty calculations. Calling $\textbf{u}$ the set of active points, the DTC predictive distribution is:
\begin{equation}
\varepsilon_{\text{DTC}}(\textbf{q}_{*}|\mathcal{D})
=\mathcal{N}(\sigma^{-2}K_{\textbf{q}_{*}\textbf{u}}\Sigma K_{\textbf{ud}}~\textbf{y},K_{\textbf{q}_{*} \textbf{q}_{*}}-Q_{\textbf{q}_{*} \textbf{q}_{*}}+K_{\textbf{q}_{*}\textbf{u}}\Sigma K_{\textbf{u}\textbf{q}_{*}}) ~~~,
\label{eq:predict_dtc}
\end{equation}
where $K$ is the Gram matrix, $\textbf{d}$ the complete dataset and:
\begin{flalign}
\Sigma&=(\sigma^{-2}K_{\textbf{ud}}K_{\textbf{du}}+K_{\textbf{uu}})^{-1} ~~~, \\
Q_{\textbf{ab}}&= K_{\textbf{au}}K_{\textbf{uu}}^{-1}K_{\textbf{ub}} ~~~.
\end{flalign}

The computationally intensive part in Eq. \ref{eq:predict_dtc} is computing the $\Sigma$ matrix, whose cost scales as $O(NM^2)$. Predictions of means and variances will then scale respectively as $O(M)$ and $O(M^2)$.

In our case, we used the ACE, Atomic Cluster Environment descriptors, proposed by Drautz \cite{Drautz2019_ACE}. The ACE expansion is based on the projection of the bonds forming the atomic environment onto a complete, orthogonal basis:
\begin{equation}
c_{inlm}=\sum_{j\in\boldsymbol{\rho}_{i}} R_{n}(r_{ij}) Y_{lm}(\hat{\textbf{r}}) f_{cut}(r_{ij}) ~~~,
\end{equation}
where $R_{n}$ is a radial basis,  $Y_{lm}$ are the well-known spherical harmonics, while $f_cut$ is a smooth cutoff function. These quantities are already permutationally invariant and can then be contracted into rotationally invariant 3-body descriptors:
\begin{equation}    
q_{in_{1}n_{2}l}^{(3)}=\sum_{m=-l}^{m=l}c_{in_{1}lm}c_{in_{2}l-m} ~~~.
\end{equation}
This tensor can be written as a vector describing the environment around the $i$-th atom, $\textbf{q}_i$. As a covariance function, we then chose the normalised square dot product:
\begin{equation}
    k(\boldsymbol{\rho}_i,\boldsymbol{\rho}_j)=\sigma^2\left( \frac{\textbf{q}_i \cdot \textbf{q}_j}{q_i q_j} \right)^2 ~~~,
\label{eq:norm_dot}\end{equation}
where $\sigma$ is a parameter roughly accounting for the variety of local energies in the dataset.

Depending on the implementation of GPR, some parameters can be trained in order to maximise the log marginal likelihood:
\begin{equation}
\text{log}~p(\textbf{y}|X)=-\frac{1}{2}\textbf{y}^{T}C^{-1}\textbf{y}-\frac{1}{2}\text{log}|C|-\frac{n}{2}\text{log}2\pi
\label{eq:marg_like}\end{equation}
Where $n$ is the number of training labels and $\textbf{y}$ and $X$ are the training labels and inputs, respectively. The marginalised likelihood measures how well a model can reflect the training set, and its local minima correspond to forms that achieve a good balance between accuracy and complexity\cite{Bauer2017_SGP}. 
\begin{figure}[h!]
    \centering
    \includegraphics[width=0.8\linewidth]{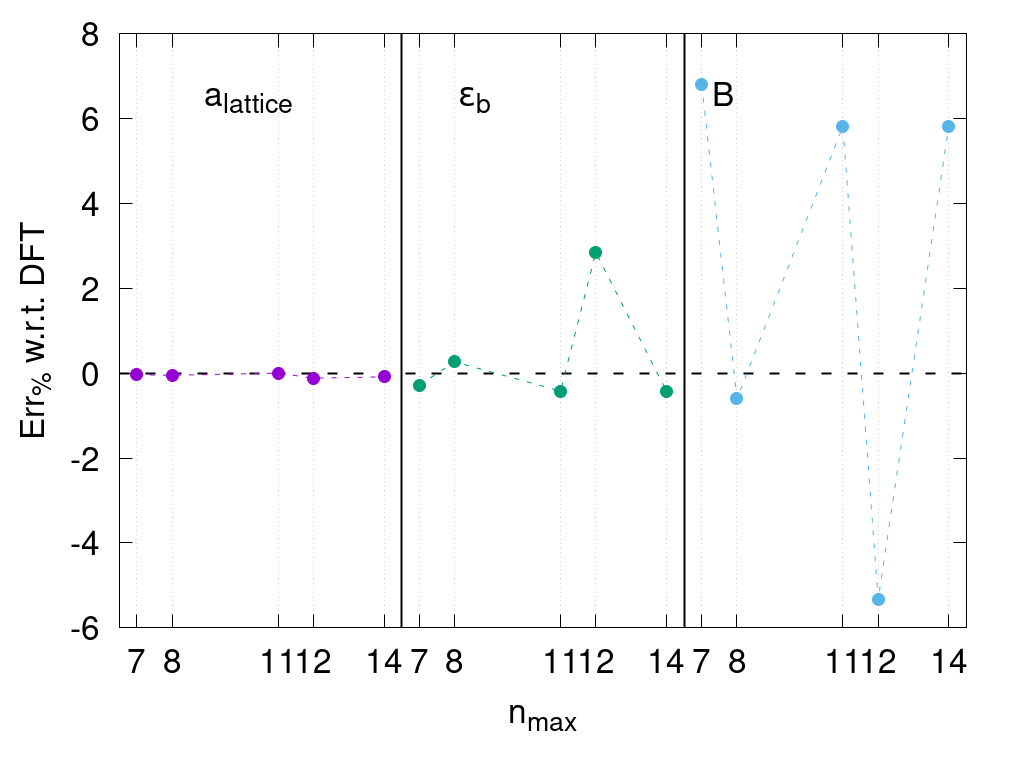}
    \includegraphics[width=0.8\linewidth]{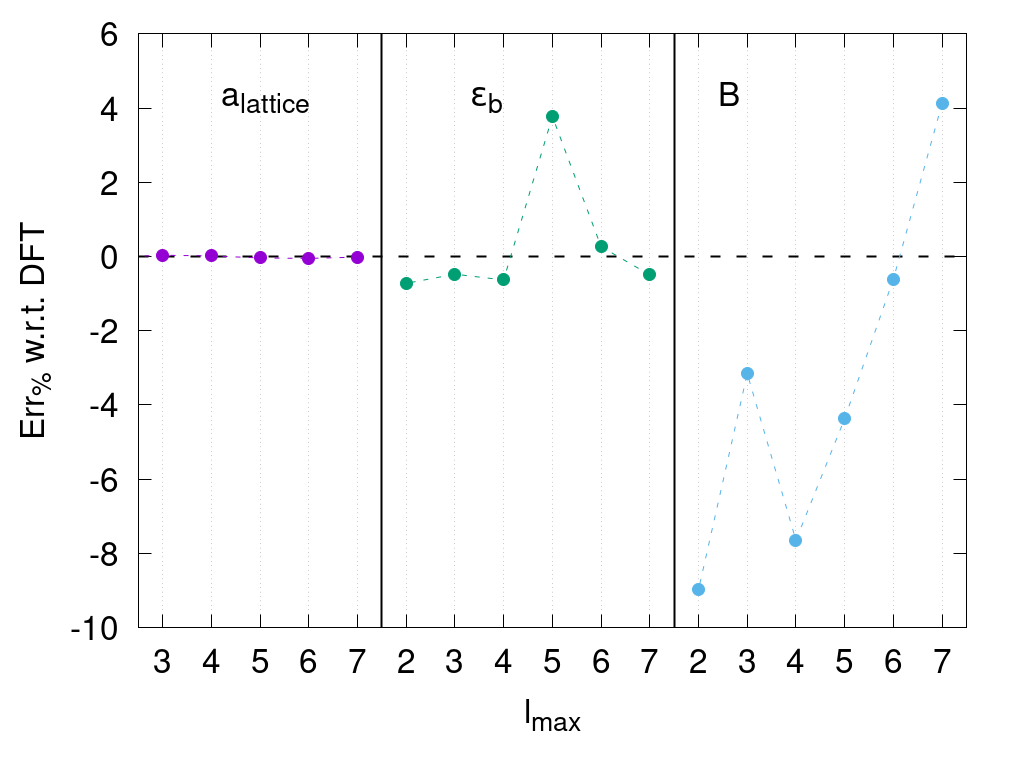}
    \includegraphics[width=0.8\linewidth]{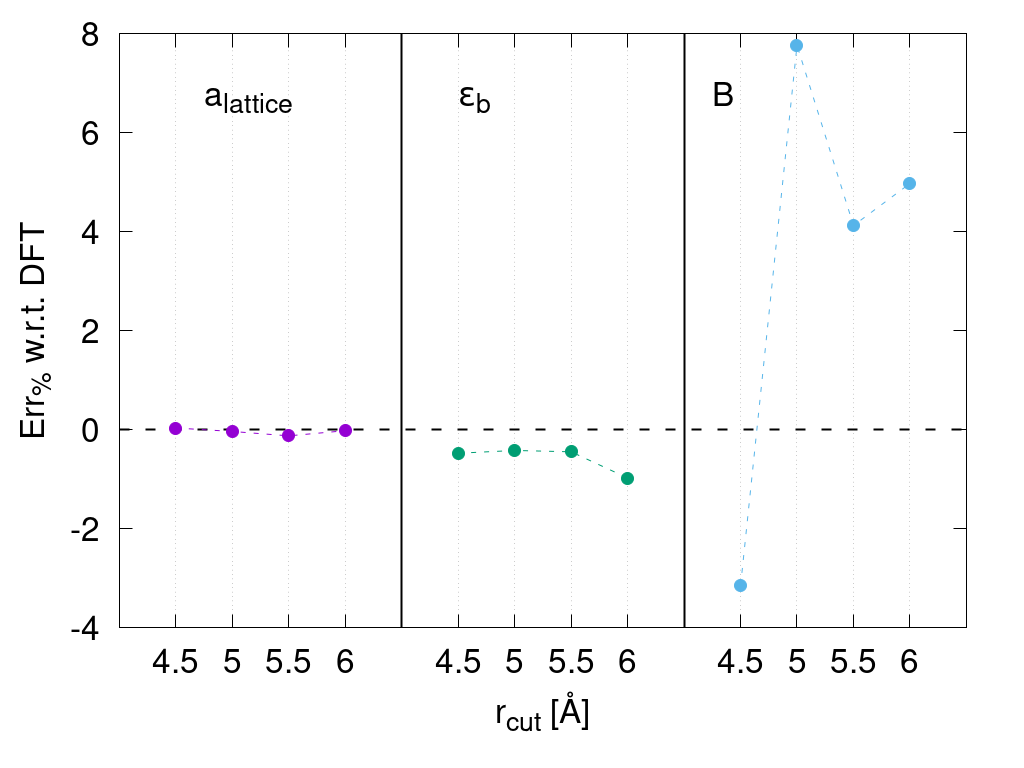}
    \caption{Relative error, relative to DFT values, for bulk properties prediction with different model parameters}
\end{figure}
\begin{figure}
    \centering
    \includegraphics[width=0.45\linewidth]{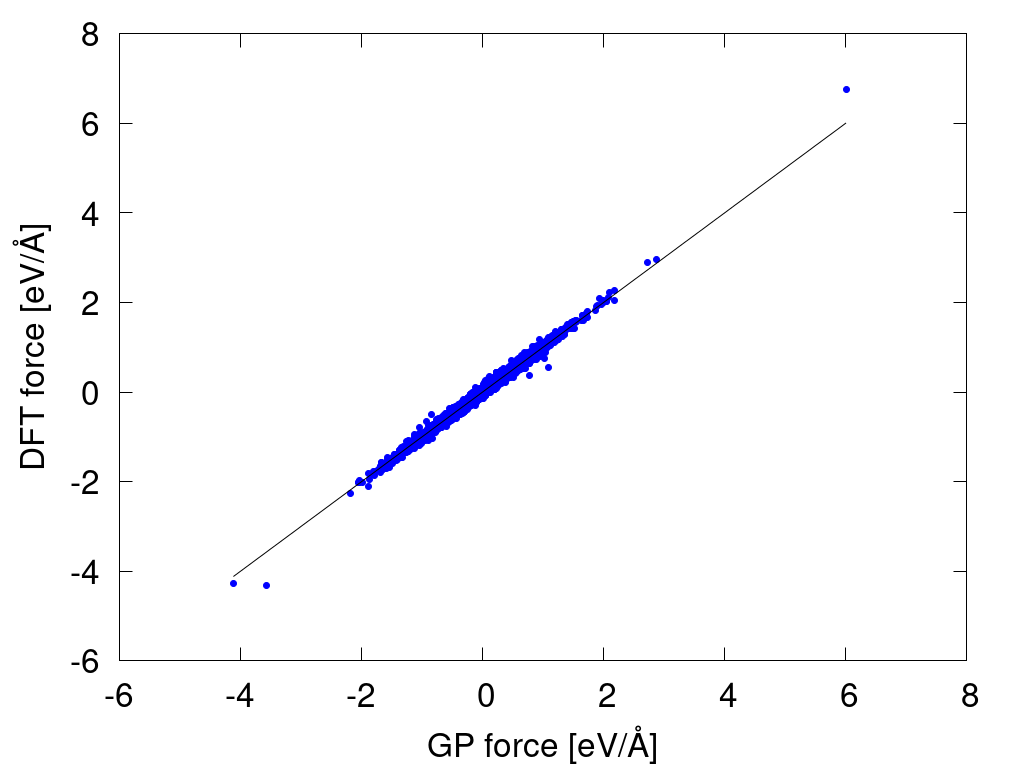}
    \includegraphics[width=0.45\linewidth]{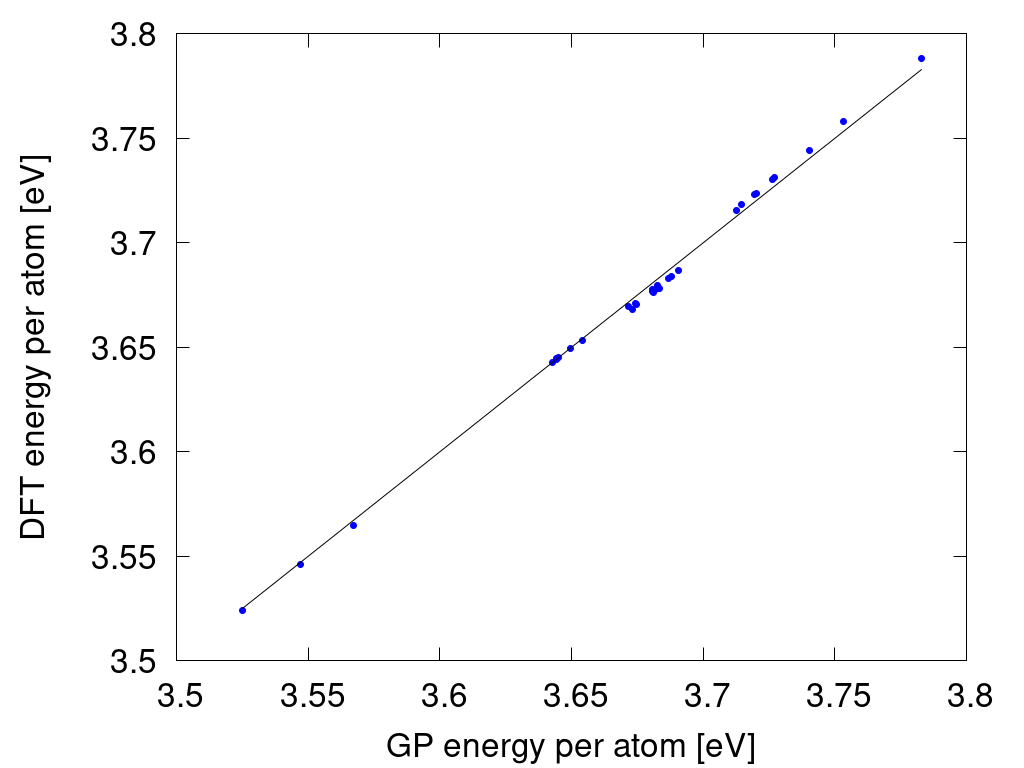}
    \caption{Parity plots for energies and forces for Al-BFF1}
    \label{fig:enter-label}
\end{figure}
\begin{figure}
    \centering
    \includegraphics[width=0.45\linewidth]{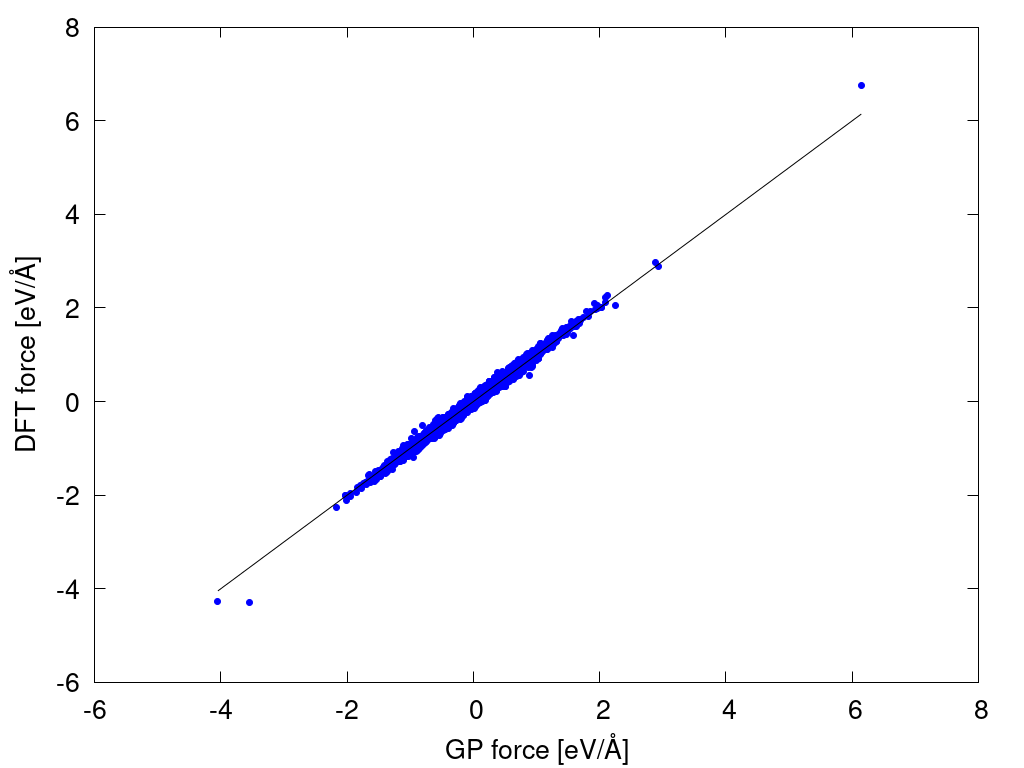}
    \includegraphics[width=0.45\linewidth]{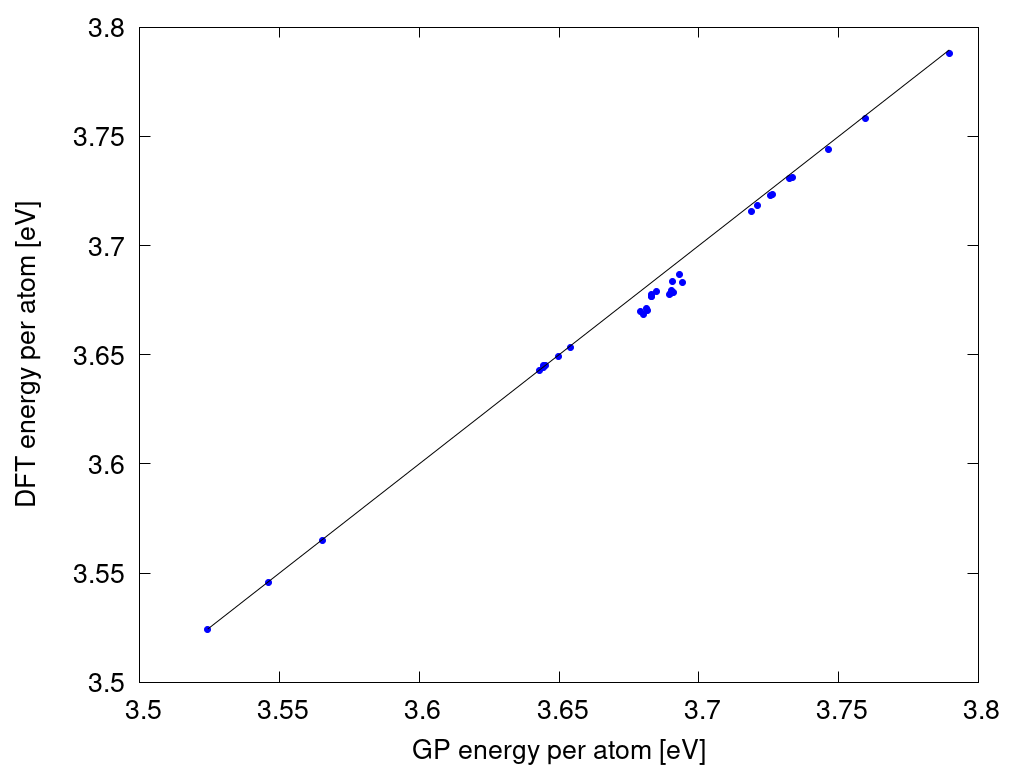}
    \caption{Parity plots for energies and forces for Al-BFF2}
    \label{fig:enter-label}
\end{figure}

\begin{figure}
\centering
    \includegraphics[width=0.45\linewidth]{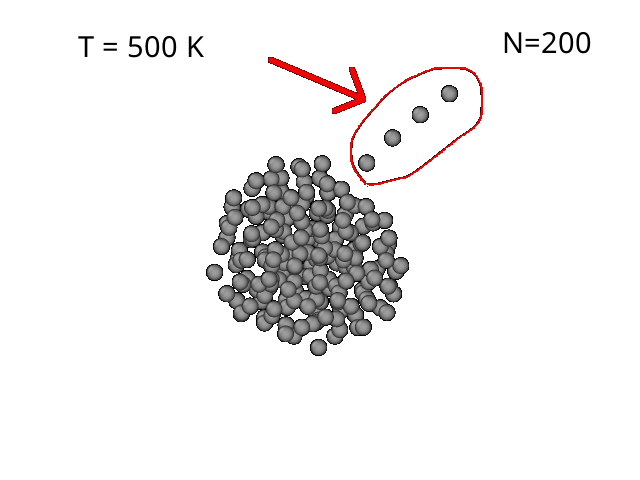}
    \includegraphics[width=0.45\linewidth]{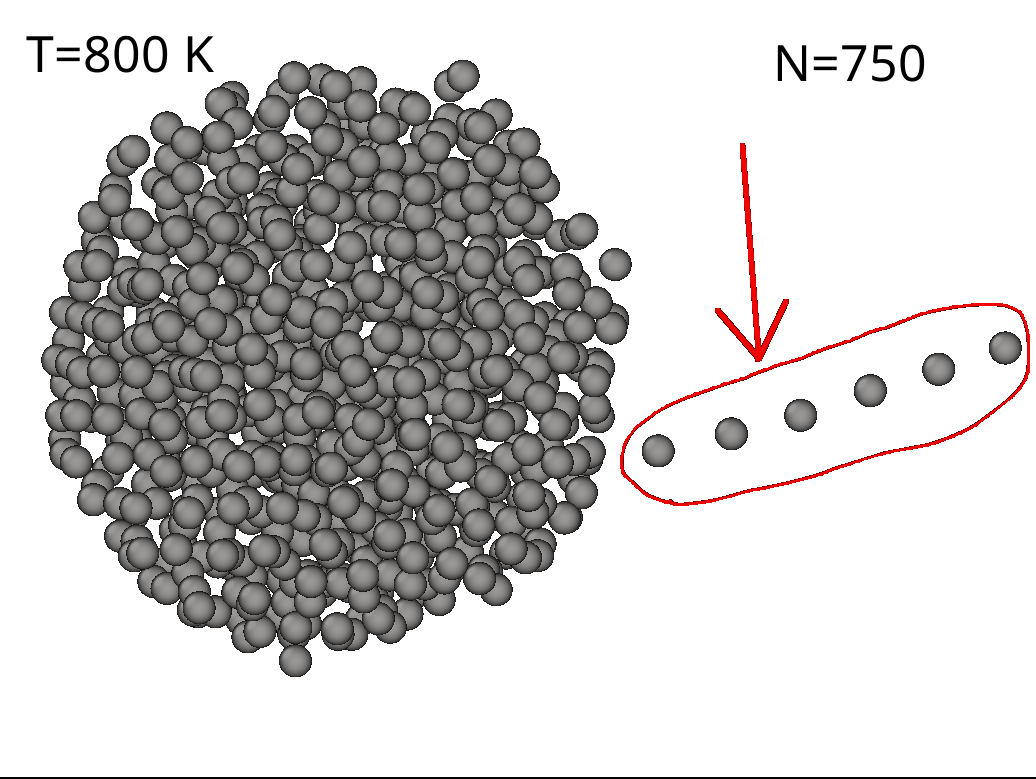}
    \caption{Examples of non-physical configurations obtained using Al-BFF 1, presenting a "hair" of atoms on the nanoparticles, which then dissociate from it, leaving the simulation cell. This behaviour is captured in two different nanoparticles and temperatures.}
    \label{fig:evapor_bff1}
\end{figure}

\begin{figure}
    \centering
    \includegraphics[width=\linewidth]{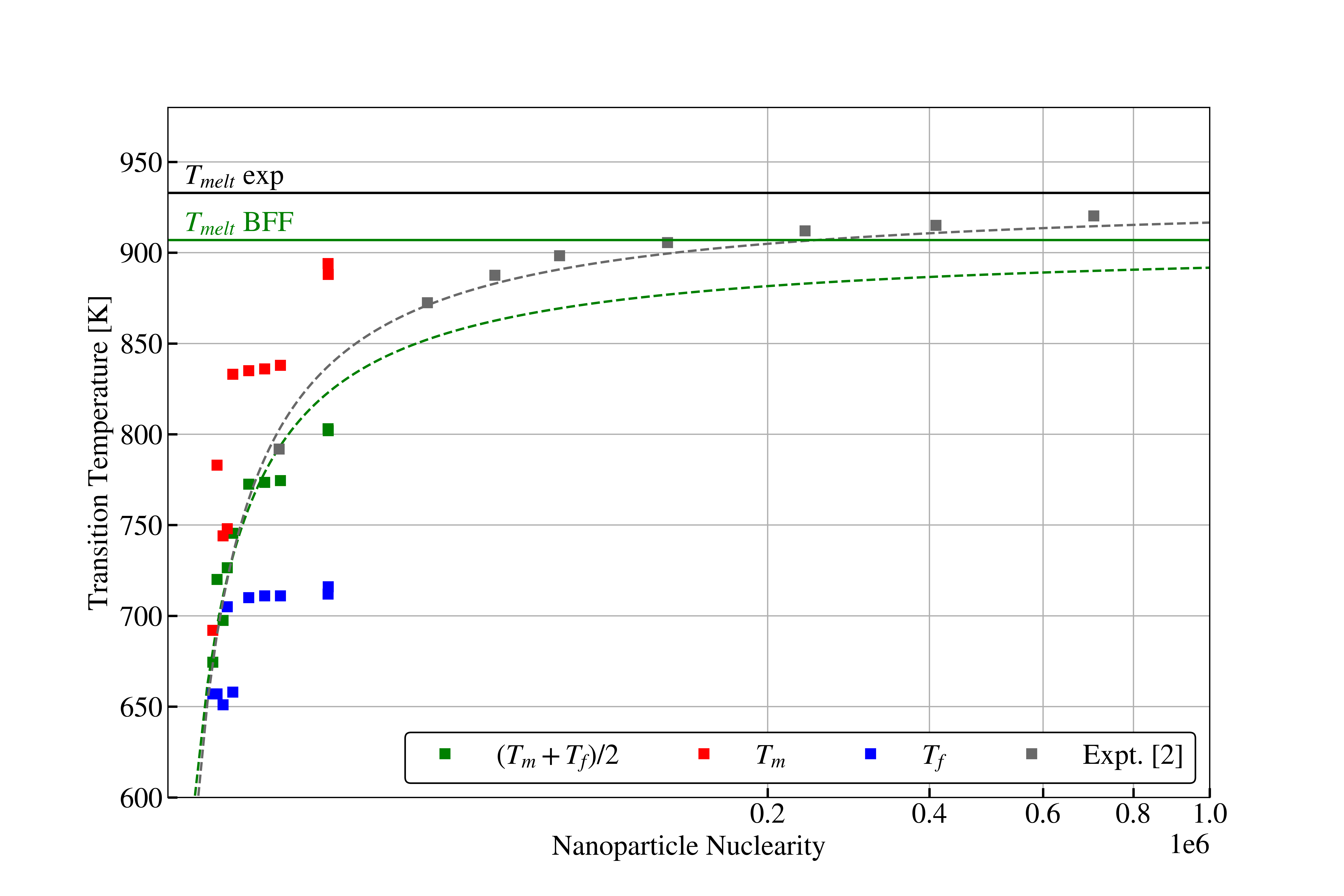}
    \caption{Melting (red) and freezing (blue) temperatures of AlNPs and their average (green) as a function of their size, compared with experimental results from  Lai \cite{lai1998melting} grey points. Dashed lines are fit to the Gibbs-Thomson equation.}
    \label{fig:fit.temps}
\end{figure}

While these comparisons can indicate whether the training procedure has been successful and allows the selection of an adequate model, they can not guarantee that the resulting trajectory, obtained in an interpolative regime, reflects the target models. The issue of the so-called long-term stability and physicality of MLPs has arisen lately following a rapid development of the field. Physicality is hard to define objectively, relying on human intuition to determine what is expected in circumstances. Sublimation was observed when simulating clusters with Al-BFF 1, with atoms detaching from the nanoparticle and leaving the simulation box below or close to the melting temperature. This behaviour can be classified as non-physical since, on the timescales of Molecular Dynamics, sublimation is practically impossible according to classical thermodynamical models. Examples of non-physical behaviour encountered are depicted in Figs. \ref{fig:evapor_bff1} a) and b).
To remediate this, simulations near the melting temperature of random clusters of size 100 and 150 were performed, and some configurations occurring near the timestep of the anomalous behaviour were hand-picked to be collected with their DFT forces and energies in the training set.
Apart from Ih, excess energies were computed for regular Dh ($n=1$ and $p=0$), MDh with either $m=n=p$ or $m=n$ and $p=m/2$, TOh with $n_{cut}=1$, COh and Wulff polyhedra. Nomenclature of indices is taken as in Baletto\cite{balettoreview}.
For every size considered, we include a snapshot of a sample structure at the beginning and at the end of the thermodynamical cycle, as well as at the highest temperature, and the plots of $p^{(1)}$ and $p^{(2)}$ at the corresponding temperatures. The phase transition temperature for all sizes is reported in Fig. \ref{fig:fit.temps}
\begin{figure}
    \centering
    \includegraphics[width=\linewidth]{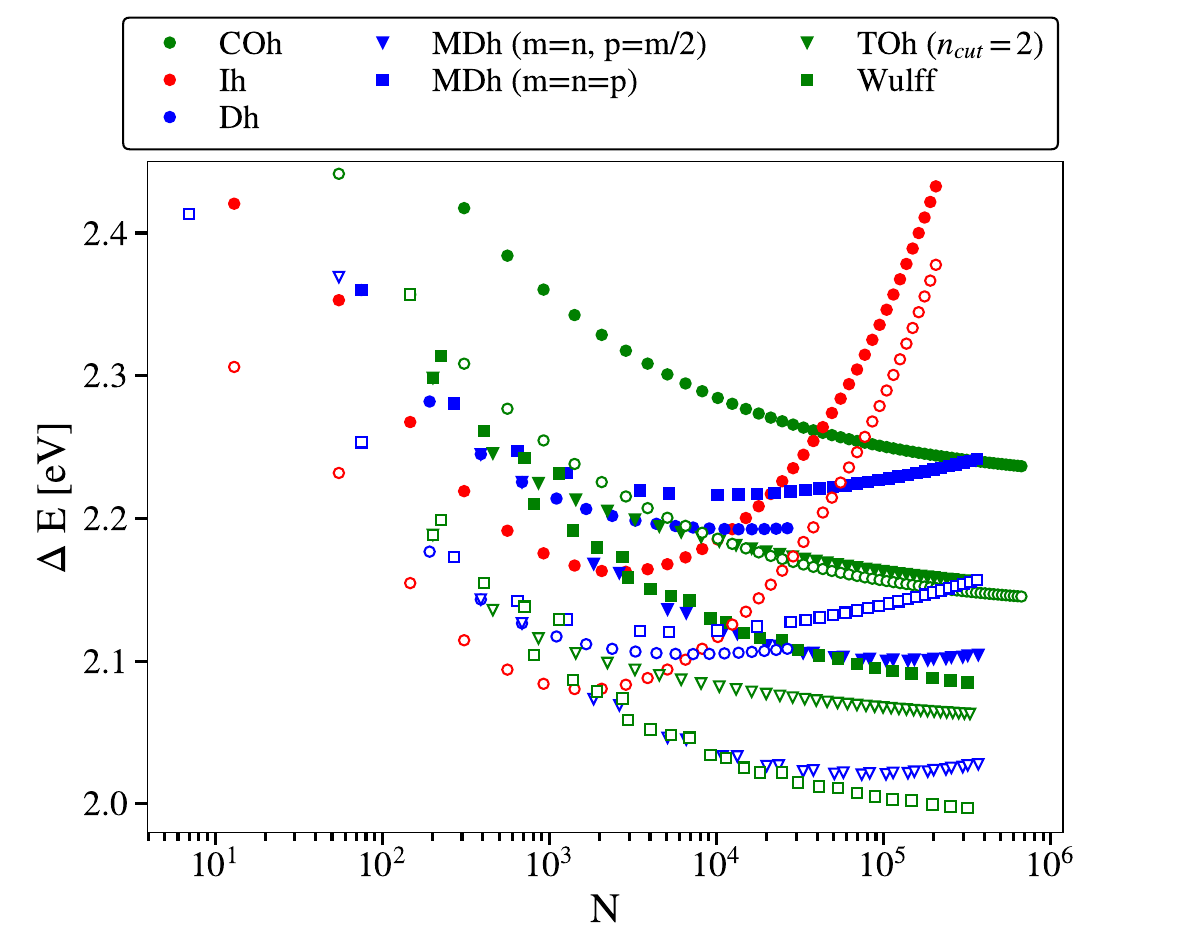}
    \caption{Excess energies of more geometrical motifs}
    \label{fig:deltas-more}
\end{figure}
\newpage
\foreach \couple in {{dh257, ih309}, {dh393, to459}, {ih561, ih923}, {ih1415, ih2057}, {ih5083, dh5096}, {to5341, ih10179}}{
\begin{figure*}
\centering
\foreach \struct in \couple{
\parbox{0.49\linewidth}{
\includegraphics[width=0.49\textwidth]{graphs/dfs/\struct.dfs.pdf}
}
\caption{Snapshots taken at the beginning of the heating, liquid droplet, and the cooling process. On the right panel, averaged PDDF and RDF for the considered AlNPs, Al{$_{\struct}$}.}
}
\end{figure*}
}

\foreach \couple in {{257, 309}, {393, 459}, {561, 923}, {1415, 2057}, {5083, 5096}, {5341, 10179}}{
\begin{figure*}
\centering
\foreach \struct in \couple{
\parbox{0.49\linewidth}{
\includegraphics[width=0.49\textwidth]{graphs/4labels/\struct.labels.pdf}
}
\caption{Atomic environments, IH (inner solid, blue), IL (inner liquid, orange), SH (surface solid, green) and SL (surface liquid, red)  as a function of temperature for the various Al{$_{\struct}$} considered. During heating, we note the appearance of solid-solid structural changes characterised by changes in the inner coordination.}
}
\end{figure*}
}

\end{document}